\documentclass[aps,prb,reprint,showpacs]{revtex4-1}

\usepackage{graphicx}
\begin{document}

\title{Ferroelectric phenomena in CdSnO$_3$: a first-principles study}

\author{A. I. Lebedev}
\email[]{swan@scon155.phys.msu.ru}
%\thanks{}
\affiliation{Physics Department, Moscow State University, Moscow, 119991 Russia}

\date{\today}

\begin{abstract}
The phonon spectrum of cubic cadmium metastannate and the crystal structures of
its distorted phases were calculated from first principles within the density
functional theory. It is shown that the phonon spectrum and the energy spectrum
of distorted phases in $\alpha$-CdSnO$_3$ are surprisingly similar to the
corresponding spectra of CdTiO$_3$. The ground state of $\alpha$-CdSnO$_3$ is
the ferroelectric $Pbn2_1$ phase; the energy gain accompanying the phase
transition from the nonpolar $Pbnm$ phase to this phase is $\sim$30~meV and
the spontaneous polarization in it is 0.25~C/m$^2$. An analysis of the eigenvector
of the ferroelectric mode in $\alpha$-CdSnO$_3$ and calculations of the partial
densities of states indicates that the ferroelectric instability in this crystal,
which does not contain $d$~transition elements, is associated with the formation
of a covalent bonding between Cd and O atoms.

\texttt{DOI: 10.1134/S1063783409090182}
\end{abstract}

%\keywords{}

\maketitle

\section{Introduction}
\label{Sec1}

In a large family of ferroelectrics with the \emph{AB}O$_3$ perovskite structure,
crystals in which $A$ is the cadmium atom are least studied. In the CdO--SnO$_2$
system there are two compounds, CdSnO$_3$ and Cd$_2$SnO$_4$.~\cite{ActaCryst.13.749}
Cadmium metastannate CdSnO$_3$ exists in two stable modifications: one with an
orthorhombically distorted perovskite structure
($\alpha$-modification)~\cite{ActaCryst.13.749,Naturwiss.31.202,JPhysChemSolids.38.877}
and the other with the rhombohedral ilmenite structure
($\beta$-modification).~\cite{JPhysChemSolids.38.877,CRAcadSci.258.3036}  Like
cadmium titanate, CdSnO$_3$ samples with the ilmenite structure can be synthesized
at temperatures $\le$800$^\circ$C~\cite{JMaterSciLett.13.1647,AnalChimActa.527.21}
and the samples with the perovskite structure at
1000--1100$^\circ$C~\cite{ActaCryst.13.749,JMaterSciLett.13.1647} or at high
pressures.~\cite{JPhysChemSolids.38.877}  In addition, CdSnO$_3$ samples with a
metastable spinel structure can be obtained upon decomposition of
CdSn(OH)$_6$.~\cite{LB-7d1g}  Cadmium orthostannate Cd$_2$SnO$_4$ also exists in
two crystal modifications: one with the orthorhombic Sr$_2$PbO$_4$ structure and
the other with the spinel structure.~\cite{LB-7d1g}  All these compounds are
$n$-type semiconductors with a band gap of 2--3~eV.~\cite{JMaterSciLett.2.505,Ferroelectrics.214.177}
Because of a high concentration of native defects, they are characterized by a rather
high conductivity ($\sigma = 10^{-5}$--$5 \times 10^3$~$\Omega^{-1}$cm$^{-1}$).
This conductivity prevents the study of these materials by dielectric methods.
Cadmium stannates are used for fabrication of conducting thin-film coatings
transparent to visible light (as transparent conductive oxides) and as gas sensors.

Cadmium metastannate with the perovskite structure is of interest as a potential
ferroelectric. Unfortunately, there have been only a few studies of this material.
The crystal structure of $\alpha$-CdSnO$_3$ was studied at 300~K on
powders~\cite{ActaCryst.13.749} and single crystals~\cite{JPhysChemSolids.38.877}
and was identified as a structure with the $Pbnm$ space group. A refined analysis
of X-ray reflection intensities~\cite{Lebedev1977} suggested the possibility of a
polar character of this structure (the proposed space group is $Pbn2_1$). The same
conclusion was made in Ref.~\onlinecite{Prutsakova2004}. An analysis of the
optical absorption and luminescence spectra~\cite{Ferroelectrics.214.177} revealed,
in the temperature dependence of the band gap $E_g(T)$ for $\alpha$-CdSnO$_3$
single crystals of unknown orientation, a rapid change in $E_g$ (by 0.11~eV) at
$T \approx 80$$^\circ$C and two more regions of change near 140 and 200$^\circ$C,
which were associated with phase transitions. In this temperature range, drastic
changes in the luminescence intensity and degree of its polarization were also
observed. If these peculiarities are really due to the ferroelectric phase
transition, we deal with a rare case when the ferroelectric properties appear in
perovskite crystals, which do not contain $d$~transition elements.

The objective difficulties of studying cadmium metastannate, incompleteness of the
experimental data in the literature, and the lack of understanding of the nature
of ferroelectricity proposed in this compound make it desirable to calculate the
physical properties of $\alpha$-CdSnO$_3$ from first principles.

\section{Calculation technique}

The calculations were performed within the density functional theory using the
pseudopotentials and plane wave expansion of wave functions as implemented in
the \texttt{ABINIT} code.~\cite{abinit}  The exchange-correlation interaction
was described in the local density approximation (LDA) following
Ref.~\onlinecite{PhysRevB.23.5048}. The pseudopotentials used were optimized
separable nonlocal pseudopotentials~\cite{PhysRevB.41.1227} constructed using
the \texttt{OPIUM} program; the local potential correction~\cite{PhysRevB.59.12471}
was added to them to improve their transferability. The parameters used for
constructing the pseudopotentials, the results of their testing, and other details
of calculations are given in Ref.~\onlinecite{PhysSolidState.51.362}.

\section{Results}

\subsection{Comparison of the properties of CdSnO$_3$ and CdTiO$_3$}

\begin{figure}
\includegraphics[scale=0.8]{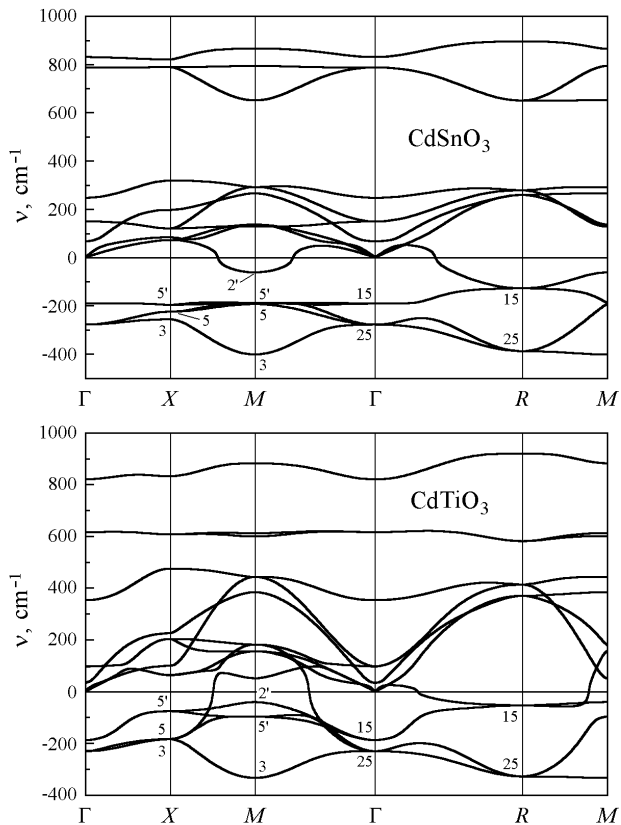}
\caption{Comparison of the phonon spectrum for the cubic parent phase of
CdSnO$_3$ with the phonon spectrum of the same phase of CdTiO$_3$
(Ref.~\onlinecite{PhysSolidState.51.362}). Labels near the curves indicate
the symmetry of the modes.}
\label{fig1}
\end{figure}

Fig.~\ref{fig1} shows the phonon spectra of $\alpha$-CdSnO$_3$ and CdTiO$_3$
crystals in the cubic parent phase with the perovskite structure. The comparison
of these spectra reveals their surprising similarity. This enables one to
suppose that other physical properties of these crystals are also similar.

\begin{table*}
\caption{\label{table1}Comparison of the relative energies of different low-symmetry
phases of CdTiO$_3$ and CdSnO$_3$. The energy of the cubic phase was taken as
the energy origin. The energies of the ground states are in boldface.}
\begin{ruledtabular}
\begin{tabular}{cccccc}
\multicolumn{3}{c}{CdTiO$_3$$^a$}          & \multicolumn{3}{c}{CdSnO$_3$} \\
\hline
Unstable mode & Space group & Energy (meV) & Unstable mode & Space group & Energy (meV) \\
\hline
$R_{15}$      & $I4/mmm$      & $-$24   & $M_5$         & $Cmma$        & $-$141 \\
$X_3$         & $P4_2/mmc$    & $-$45   & $M_5$         & $Pmna$        & $-$262 \\
$\Gamma_{25}$ & $P{\bar 4}m2$ & $-$134  & $X_3$         & $P4_2/mmc$    & $-$404 \\
$X_5$         & $Pmma$        & $-$160  & $X_5$         & $Pmma$        & $-$474 \\
$\Gamma_{15}$ & $R3m$         & $-$245  & $R_{15}$      & $I4/mmm$      & $-$536 \\
$X_5$         & $Cmcm$        & $-$282  & $\Gamma_{25}$ & $P{\bar 4}m2$ & $-$540 \\
$\Gamma_{15}$ & $P4mm$        & $-$340  & $\Gamma_{15}$ & $R3m$         & $-$753 \\
$\Gamma_{15}$, $\Gamma_{25}$ & $Amm2$ & $-$412 & $X_5$  & $Cmcm$        & $-$886 \\
$\Gamma_{25}$ & $R32$         & $-$486  & $\Gamma_{15}$ & $P4mm$        & $-$1079 \\
$R_{25}$      & $I4/mcm$      & $-$912  & $\Gamma_{15}$, $\Gamma_{25}$ & $Amm2$ & $-$1259 \\
$M_3$         & $P4/mbm$      & $-$920  & $\Gamma_{25}$ & $R32$         & $-$1450 \\
$R_{25}$      & $R{\bar 3}c$  & $-$1197 & $M_3$         & $P4/mbm$      & $-$1617 \\
$R_{15}$      & $C2/m$        & $-$1202 & $R_{25}$      & $I4/mcm$      & $-$1659 \\
$R_{25} + M_3$ & $Pbnm$       & $-$1283 & $R_{15}$      & $C2/m$        & $-$2455 \\
$B_{2u}$      & $Pb2_1m$      & $-$1285 & $R_{25}$      & $R{\bar 3}c$  & $-$2460 \\
$B_{1u}$      & $Pbn2_1$      & {\bf $-$1290} & $R_{25} + M_3$ & $Pbnm$       & $-$2575 \\
              &               &         & $B_{1u}$      & $Pbn2_1$      & {\bf $-$2605} \\
\end{tabular}
\end{ruledtabular}
{\footnotesize $^a$References~\onlinecite{PhysSolidState.51.362,PhysSolidState.51.802}. \hfill}
\end{table*}

The energies of different low-symmetry phases formed upon distortion of the cubic
parent phase of CdSnO$_3$ according to the eigenvectors of unstable phonons found
in its phonon spectrum are given in Table~\ref{table1}. For comparison, the results
of previous calculations of the energies of distorted phases in cadmium
titanate~\cite{PhysSolidState.51.362,PhysSolidState.51.802}  are included in the
table. The existence of one more unstable $M_5$ mode in CdSnO$_3$, which describes
the octahedral rotations in the structure, results in two more low-symmetry phases
with $Cmma$ and $Pmna$ space groups. It is evident that the energy spectra of
different distorted phases in $\alpha$-CdSnO$_3$ and CdTiO$_3$ are very similar,
which confirms the similarity of these crystals. The nonpolar phase with the lowest
energy in cadmium metastannate is the orthorhombic $Pbnm$ phase.

\begin{table*}
\caption{\label{table2}Lattice parameters $a$, $b$, and $c$ (in {\AA}) and atomic
coordinates in CdSnO$_3$ crystals with $Pbnm$ and $Pbn2_1$ space groups.}
\begin{ruledtabular}
\begin{tabular}{ccccccc}
Parameter & \multicolumn{2}{c}{This work}   & \multicolumn{4}{c}{Experiment} \\
\cline{2-7}
          & $Pbnm$     & $Pbn2_1$   & Ref.~\onlinecite{ActaCryst.13.749} & Ref.~\onlinecite{JPhysChemSolids.38.877} & Ref.~\onlinecite{Prutsakova2004} & Ref.~\onlinecite{InorgChem.43.1667}$^a$ \\
\hline
$a$       & 5.5024     & 5.5284     & 5.547 & 5.4578 & 5.4593 & 5.4588 \\
$b$       & 5.5982     & 5.5972     & 5.577 & 5.5773 & 5.5804 & 5.5752 \\
$c$       & 7.9770     & 7.9584     & 7.867 & 7.8741 & 7.8771 & 7.8711 \\
Cd$_x$    & $-$0.00861 & $-$0.00553 &       &        &        & $-$0.0092 \\
Cd$_y$    & +0.04816   & +0.03818   &       &        &        & +0.0423 \\
Cd$_z$    & +0.25000   & +0.26021   &       &        &        & +0.2500 \\
Sn$_x$    & +0.00000   & +0.00036   &       &        &        & +0.0000 \\
Sn$_y$    & +0.50000   & +0.51455   &       &        &        & +0.5000 \\
Sn$_z$    & +0.00000   & $-$0.00209 &       &        &        & +0.0000 \\
O1$_x$    & +0.12498   & +0.12340   &       &        &        & +0.114 \\
O1$_y$    & +0.42809   & +0.42802   &       &        &        & +0.455 \\
O1$_z$    & +0.25000   & +0.24122   &       &        &        & +0.250 \\
O2$a_x$   & +0.68331   & +0.66193   &       &        &        & +0.695 \\
O2$a_y$   & +0.31304   & +0.35381   &       &        &        & +0.301 \\
O2$a_z$   & +0.06830   & +0.04275   &       &        &        & +0.058 \\
O2$b_x$   & +0.31669   & +0.29516   &       &        &        & +0.305 \\
O2$b_y$   & +0.68696   & +0.72657   &       &        &        & +0.699 \\
O2$b_z$   & $-$0.06830 & $-$0.09020 &       &        &        & $-$0.058 \\
\end{tabular}
\end{ruledtabular}
{\footnotesize $^a$In the structure determination, the space group was assumed to be $Pbnm$. \hfill}
\end{table*}

Table~\ref{table2} presents the calculated structural parameters for the $Pbnm$
phase of cadmium metastannate. As follows from the comparison with the literature
data, the calculated lattice parameters are close to the experimentally observed
ones. A slight overestimate of the calculated lattice parameters is associated
with the peculiarity of the Sn pseudopotential, which manifests itself in tests
as overestimated lattice parameters of SnO$_2$ and gray tin. The atomic
coordinates in the unit cell agree well with the results of the structure
determination of $\alpha$-CdSnO$_3$,~\cite{InorgChem.43.1667} which was
performed under the assumption that the space group of the crystal is $Pbnm$.

Simultaneously with calculating the structure of $\alpha$-CdSnO$_3$, the structure
of $\beta$-CdSnO$_3$ (space group $R{\bar 3}$) was obtained. The calculated
lattice parameters of this phase ($a = 5.5238$~{\AA}, $c = 14.6550$~{\AA}) were
found to be close to those obtained experimentally ($a = 5.4530$~{\AA},
$c = 14.960$~{\AA}, Ref.~\onlinecite{JPhysChemSolids.38.877}). The energy of
the $R{\bar 3}$ phase of cadmium metastannate was by 77~meV lower than that
of the $Pbn2_1$ phase. This means that at $T = 0$ the phase with the distorted
perovskite structure is metastable. It should be noted that according to our
data, in CdTiO$_3$, which also exists in ilmenite and perovskite modifications,
the energy of the $R{\bar 3}$ phase at $T = 0$ is by 166~meV lower than that
of the $Pbnm$ phase. The difference in the entropy contributions to the
thermodynamic potential $\Phi$ can result in intersection of the curves for
$\Phi(T)$ for the two considered phases. This can explain why one obtains
crystals with the ilmenite structure at a low synthesis temperature and crystals
with the perovskite structure at a high synthesis temperature (see Sec.~\ref{Sec1}).

\subsection{Ferroelectric phase in CdSnO$_3$}

In order to check the possibility of ferroelectricity in $\alpha$-CdSnO$_3$,
the frequencies of phonons at the $\Gamma$ point were calculated for the
orthorhombic $Pbnm$ phase of this crystal. The calculations revealed one
unstable $B_{1u}$ mode with a frequency of 89$i$~cm$^{-1}$, which can be
associated with the ferroelectric phase transition $Pbnm \to Pbn2_1$.

The calculated equilibrium atomic positions and lattice parameters for the
$Pbn2_1$ phase are given in Table~\ref{table2}. The lattice distortion is
accompanied by a noticeable rearrangement of the local environment of the Cd
atom; it results in that the average distance to four nearest oxygen atoms
increases by 0.017~{\AA} but another two oxygen atoms become closer by 0.23~{\AA}.
Thus, the ferroelectric lattice distortion is accompanied by an increase in the
effective coordination number of the Cd atom. The energy gain from the
$Pbnm \to Pbn2_1$ phase transition is $\Delta E = 30.5$~meV (Table~\ref{table1}).
The sufficiently high energy gain enables one to expect that the structure will
be ferroelectric at room temperature, in agreement with the data of
Refs.~\onlinecite{Lebedev1977,Prutsakova2004}. Indeed, the phase transition
temperature ($\Delta E /k \sim 350$~K) estimated from the energy gain upon
the transition to the ferroelectric phase is close to the temperature of
80$^\circ$C, at which the most drastic changes in the absorption spectra were
observed.~\cite{Ferroelectrics.214.177}  The calculation of the spontaneous
polarization in CdSnO$_3$ with $Pbn2_1$ structure using the Berry phase
method~\cite{PhysRevB.66.104108}  gives an unexpectedly high value of
$P_s = 0.25$~C/m$^2$, which is close to spontaneous polarization in barium
titanate.

\section{Discussion}

The results of our calculations show that the ferroelectricity can appear in
perovskites that do not contain atoms of $d$ transition elements. Let us
try to understand the nature of such phase transitions.

The appearance of the ferroelectric phase transition in $\alpha$-CdSnO$_3$
cannot be associated with an off-centering of cadmium atoms. Although in
the cubic parent phase of CdSnO$_3$ the diagonal element of the on-site
force constant matrix for Cd atoms is $-$0.0111~Ha/Bohr$^2$, which indicates
its off-center position, in the $Pbnm$ phase, after the unit cell volume was
decreased by 8.4\%, the minimum value of the diagonal element increases to
+0.0714~Ha/Bohr$^2$, and the potential well for the Cd atom becomes on-center.%
    \footnote{In this work, the values of the force constants and the partial
    densities of states are given in Hartree atomic units.}

\begin{table*}
\caption{\label{table3}Eigenvectors $\xi$ of the ferroelectric $B_{1u}$ mode
and effective charges $Z^*$ of atoms in CdSnO$_3$ and CdTiO$_3$ crystals with
$Pbnm$ space group and in BaTiO$_3$ with $Pm3m$ space group.}
\begin{ruledtabular}
\begin{tabular}{ccccccc}
Atom & $\xi_x$ & $\xi_y$ & $\xi_z$ & $Z^*_{xx}$ & $Z^*_{yy}$ & $Z^*_{zz}$ \\
\hline
Cd   & +0.00000   & +0.00000 & +0.18913   & +2.462   & +2.506   & +2.440 \\
Sn   & +0.00257   & +0.12515 & $-$0.01935 & +4.379   & +4.407   & +4.338 \\
O1   & +0.00000   & +0.00000 & $-$0.05940 & $-$2.052 & $-$1.894 & $-$2.905 \\
O2   & $-$0.13111 & +0.20564 & $-$0.19455 & $-$2.395 & $-$2.509 & $-$1.937 \\
\hline
Cd   & +0.00000   & +0.00000   & +0.10628   & +2.570   & +2.500   & +2.592 \\
Ti   & +0.01863   & $-$0.19067 & +0.15463   & +7.363   & +7.693   & +7.260 \\
O1   & +0.00000   & +0.00000   & $-$0.16286 & $-$2.194 & $-$1.957 & $-$5.650 \\
O2   & $-$0.07161 & +0.18255   & $-$0.19322 & $-$3.869 & $-$4.118 & $-$2.101 \\
\hline
Ba   &            &            & +0.02988   &          &          & +2.738 \\
Ti   &            &            & +0.67340   &          &          & +7.761 \\
O1   &            &            & $-$0.54043 &          &          & $-$6.128 \\
O2   &            &            & $-$0.35607 &          &          & $-$2.186 \\
\end{tabular}
\end{ruledtabular}
\end{table*}

Now, we consider the characteristics of the ferroelectric soft mode. The
eigenvectors $\xi$ of the dynamic matrix for the $B_{1u}$ mode in $\alpha$-CdSnO$_3$
and CdTiO$_3$ and, for comparison, of the $\Gamma_{15}$ mode in BaTiO$_3$ are
given in Table~\ref{table3}. Here, O1 denotes the oxygen atoms which form, along
with the $B$ atoms, the chains which propagate along the polar axis of the
ferroelectric phase, and O2 denotes the other oxygen atoms. An analysis of the
eigenvector of the $B_{1u}$ mode in $\alpha$-CdSnO$_3$ shows that the main
contribution to this ferroelectric mode is made by Cd and O2 atoms. The Sn atoms
exhibit substantial displacements only in the $y$ direction perpendicular to the
polar axis and their small displacements along the polar axis are out-of-phase with
the Cd atoms and in-phase with the oxygen atoms. This means that cadmium atoms play
the main role in the appearance of ferroelectricity in $\alpha$-CdSnO$_3$. From
the comparison of the eigenvectors of ferroelectric modes in the above three
crystals it follows that CdSnO$_3$ and BaTiO$_3$ can be considered as limiting
cases in which the appearance of ferroelectricity is caused by $A$ and $B$ atoms
in the \emph{AB}O$_3$ perovskite structure, whereas cadmium titanate can be
regarded as an intermediate case.

The Born effective charges $Z^*$ of atoms in the $Pbnm$ phase of cadmium metastannate
(Table~\ref{table3}) are close to nominal charges of constituent ions. In the
cubic parent phase, the effective charge of Sn atoms is slightly smaller than that
in the orthorhombic phase ($Z^* = 4.17$) and, for Cd atoms, it is slightly larger
($Z^* = 3.26$).

\begin{figure}
\includegraphics{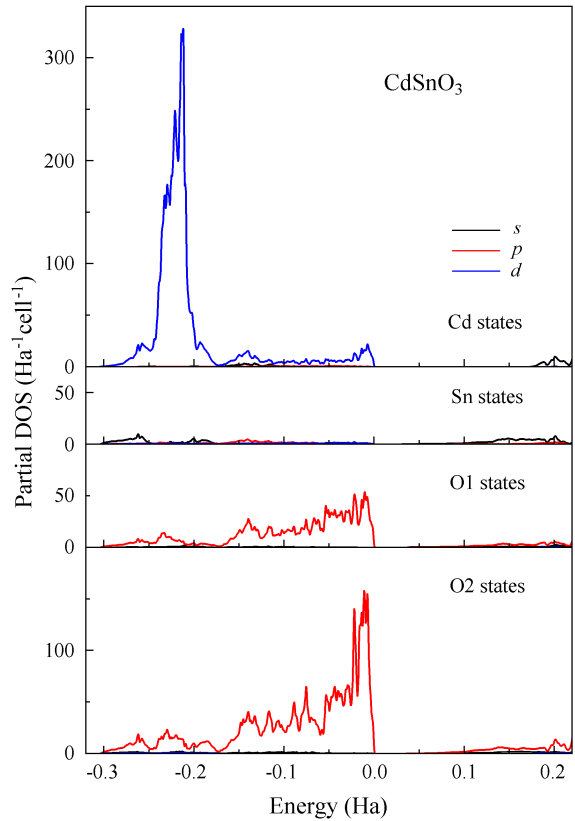}
\caption{(Color online) Partial contributions of the $s$, $p$, and $d$ states
of Cd, Sn, O1, and O2 atoms to the density of states in the $Pbnm$ phase of
CdSnO$_3$. The energy origin is taken at the top of the valence band.}
\label{fig2}
\end{figure}

For better understanding of the nature of ferroelectric instability and of the
character of chemical bonding in $\alpha$-CdSnO$_3$, the partial densities of
states (contributions of $s$, $p$, and $d$ orbitals of Cd, Sn, O1, and O2 atoms
to the total density of states) were calculated. The results of these calculations
are presented in Fig.~\ref{fig2}. As follows from the figure, the overlap of Cd $4d$ states
and O $2p$ states plays an important role in the formation of chemical bonding
in the crystal; comparable contributions from these states give evidence for
a noticeable covalent component in this bonding. The contribution of Sn $5s$ and
Sn $5p$ states to the valence band is much smaller than that of Cd atoms, which
suggests a predominantly ionic character of the Sn--O bond. The Sn $5s$ states
overlapping with O $2p$ states make the main contribution to the conduction band
($E > 0.037$~Ha in Fig.~\ref{fig2}). It should be noted that, although the partial
densities of states for $\alpha$-CdSnO$_3$ have been already calculated in
Ref.~\onlinecite{InorgChem.43.1667},  the role of Cd $4d$ states, whose contribution
to the density of states is several times greater than the contribution of
Sn $5s$ states according to our data, was not analyzed in that paper. The
conclusion about the important role of covalent interaction between Cd and O atoms,
which follows from the analysis of partial densities of states, agrees with the
results of the analysis of the eigenvector of the ferroelectric mode and suggests
that the rearrangement of these bonds is the cause of the ferroelectric instability
in $\alpha$-CdSnO$_3$.

\begin{table}
\caption{\label{table4}Calculated frequencies $\nu_i$ of the IR and Raman active
modes for CdSnO$_3$ crystals with $Pbnm$ and $Pbn2_1$ space groups.}
\begin{ruledtabular}
\begin{tabular}{ccc}
Space  & Mode     & $\nu_i$ (cm$^{-1}$) \\
group  &          & \\
\hline
$Pbnm$ & $A_g$    & 79, 130, 196, 290, 413, 452, 522 \\
       & $B_{1g}$ & 108, 150, 232, 347, 440, 500, 671 \\
       & $B_{2g}$ & 87, 207, 446, 474, 668 \\
       & $B_{3g}$ & 107, 142, 375, 500, 590 \\
       & $B_{1u}$ & 89$i$, 118, 163, 237, 390, 551, 594 \\
       & $B_{2u}$ & 62, 150, 175, 226, 264, 302, 420, 512, 605 \\
       & $B_{3u}$ & 90, 131, 186, 216, 267, 362, 420, 476, 640 \\
\hline
$Pbn2_1$ & $A_1$  & 78, 117, 138, 144, 201, 213, 241, 271, \\
         &        & 375, 413, 436, 504, 562, 574 \\
         & $A_2$  & 93, 105, 121, 130, 149, 202, 226, 263, \\
         &        & 351, 396, 432, 486, 539, 599, 655 \\
         & $B_1$  & 90, 120, 152, 206, 236, 251, 273, 351, \\
         &        & 396, 443, 466, 483, 610, 658 \\
         & $B_2$  & 88, 131, 151, 193, 208, 238, 261, 298, \\
         &        & 364, 421, 462, 515, 579, 603 \\
\end{tabular}
\end{ruledtabular}
\end{table}

Table~\ref{table4} presents the calculated frequencies of modes active in Raman
and infrared (IR) spectra for CdSnO$_3$ crystals with $Pbnm$ and $Pbn2_1$ space
groups. In a crystal with the $Pbnm$ structure, there are 24 Raman active modes
(modes with $A_g$, $B_{1g}$, $B_{2g}$, and $B_{3g}$ symmetries) and 25 IR active
modes (modes with $B_{1u}$, $B_{2u}$, and $B_{3u}$ symmetries). When the symmetry
of a crystal is lowered to $Pbn2_1$, all 57 optical modes become active in the
Raman spectra and 42 modes of $A_1$, $B_1$, and $B_2$ symmetry become active
in the IR spectra. Unfortunately, the IR absorption spectra for $\alpha$-CdSnO$_3$
obtained in the 250--800~cm$^{-1}$ range in Ref.~\onlinecite{ZAnorgAllgChem.465.186}
consist of five very wide bands, and their identification was not possible.

As for the prospects of further studies of ferroelectric properties of $\alpha$-CdSnO$_3$,
we should note the following. As was noted in Sec.~\ref{Sec1}, these crystals are $n$-type
semiconductors and their temperature dependence of conductivity follows the Arrhenius
law with an activation energy of 0.3~eV.~\cite{JPhysChemSolids.38.877} Therefore, if
the hopping conductivity is not high, studies of them at low temperatures are possible.
Donor levels supplying electrons to the conduction band are usually attributed to
the oxygen vacancies. However, a rather small effect of variations in O$_2$ partial
pressure on the conductivity of $\alpha$-CdSnO$_3$ in the annealing
experiments~\cite{JMaterSciLett.2.505} makes this explanation unlikely. In
Ref.~\onlinecite{ApplPhysLett.80.1376}, in discussing the properties of Cd$_2$SnO$_4$,
the energy positions of various defects were calculated and it was shown that the
Cd$_{\rm Sn}$ antisite defect is the most important defect in this crystal. Perhaps,
the same defects are responsible for the electrical conductivity of $\alpha$-CdSnO$_3$.
We hope that improvements in the technology of the crystal growth and doping of them
with acceptors will make it possible in the future to obtain high-resistivity crystals
of $\alpha$-CdSnO$_3$, which will enable one to perform direct dielectric measurements.

\section{Conclusions}

The first-principles calculations confirm the existence of a stable ferroelectric
$Pbn2_1$ phase in cadmium metastannate, whose energy is by 30.5~meV lower than that
of the nonpolar $Pbnm$ phase and whose spontaneous polarization is 0.25~C/m$^2$.
An analysis of the eigenvector of the ferroelectric mode and calculations of the
partial densities of states show that the ferroelectric instability in $\alpha$-CdSnO$_3$,
which does not contain $d$~transition elements, is associated with the formation of
the covalent bonding between Cd and O atoms.

\begin{acknowledgments}
This work was supported by the Russian Foundation for Basic Research (project no. 08-02-01436).
\end{acknowledgments}

%\bibliography{all}
\providecommand{\BIBYu}{Yu}

\end{document}